\documentclass[12pt,fleqn]{article}
\pdfoutput=1
\usepackage{amsmath}
\usepackage{amsthm}
\usepackage{amsthm}
\usepackage{amsfonts}
\usepackage{amssymb}
\usepackage{amscd}
\usepackage{mathtools}
\usepackage{fullpage}
\usepackage{color}
\usepackage{times}
\usepackage{graphicx}
\usepackage{float}

\begin{document}
\renewcommand{\thefootnote}{\fnsymbol{footnote}}
\title{Numerical examination of commutativity between Backus and Gazis et~al.\ averages}
\author{
David R. Dalton\footnote{
Department of Earth Sciences, Memorial University of Newfoundland,
{\tt dalton.nfld@gmail.com}},
Michael A. Slawinski \footnote{
Department of Earth Sciences, Memorial University of Newfoundland,
{\tt mslawins@mac.com}}
}
\date{September 5, 2016}
\maketitle
\renewcommand{\thefootnote}{\arabic{footnote}}
\setcounter{footnote}{0}
\section*{Abstract}
Dalton and Slawinski~(2016) show that, in general, the Backus~(1962) average and the Gazis et~al.~(1963) average do not commute.
Herein, we examine the extent of this noncommutativity.
We illustrate numerically that the extent of noncommutativity is a function of the strength of anisotropy.
The averages nearly commute in the case of weak anisotropy.
\section{Introduction}
Dalton and Slawinski~(2016) show that---in general---the Backus~(1962) average, which is an average over a spatial variable, and the Gazis et~al.\ (1963) average, which is an average over a symmetry group, do not commute.
These averages result in the so-called equivalent and effective media, respectively.
In this paper, using the monoclinic and orthotropic symmetries, we numerically study the extent of the lack of commutativity between these averages.
Also, we examine the effect of the strength of the anisotropy on noncommutativity.
We consider the following diagram.
\begin{equation}
\label{eq:CD}
\begin{CD}
\rm{mono}@>\rm{B}>>\rm{mono}\\
@V\mathrm{G}VV                         @VV\rm{G}V\\
\rm{ortho}@>>\rm{B}>\rm{ortho}
\end{CD}\,.
\end{equation}
Herein, $\rm B$ and $\rm G$ stand for the Backus~(1962) average and the Gazis et~al.\ (1963) average, respectively.
The upper left-hand corner of Diagram~\ref{eq:CD} is a series of parallel monoclinic layers.
The lower right-hand corner is a single orthotropic medium.
The intermediate clockwise result is a single monoclinic tensor: an equivalent medium; the intermediate counterclockwise result is a series of parallel orthotropic layers: effective media.

Given monoclinic tensors,~$c_{ijk\ell}$\,, in the upper left-hand corner of Diagram~\ref{eq:CD} and following the clockwise path, we have\,, according to Dalton and Slawinski~(2016) and Bos et~al.~(2016),
\begin{equation*}
c_{1212}^\circlearrowright=
\overline{c_{1212}}-\overline{\left(\frac{c_{3312}^2}{c_{3333}}\right)}+
\overline{\left(\frac{1}{c_{3333}}\right)}^{\,\,-1}
\overline{\left(\frac{c_{3312}}{c_{3333}}\right)}^{\,2}\,,
\end{equation*}
\begin{equation*}
c_{1313}^\circlearrowright=\overline{\left(\frac{c_{1313}}{D}\right)}/(2D_2)\,,\qquad
c_{2323}^\circlearrowright=\overline{\left(\frac{c_{2323}}{D}\right)}/(2D_2)\,,
\end{equation*}
where $D\equiv 2(c_{2323}c_{1313}-c_{2313}^2)$ and
$D_2\equiv (\overline{c_{1313}/D})(\overline{c_{2323}/D})-(\overline{c_{2313}/D})^2$\,; $c_{ijk\ell}^\circlearrowright$ are the elasticity parameters of the orthotropic tensor in the lower right-hand corner.
Following the counterclockwise path, we have
\begin{equation*}
c_{1212}^\circlearrowleft=\overline{c_{1212}}\,,\quad
c_{1313}^\circlearrowleft=\overline{\left(\frac{1}{c_{1313}}\right)}^{\,\,-1}\,,\quad
c_{2323}^\circlearrowleft=\overline{\left(\frac{1}{c_{2323}}\right)}^{\,\,-1}\,.
\end{equation*}
The other parameters are the same for both paths.

As stated by Dalton and Slawinski~(2016), the results of the clockwise and  counterclockwise paths are the same for all elasticity parameters if $c_{2313}=c_{3312}=0$\,, which is a special case of monoclinic symmetry.
For that case, the Backus~(1962) average and the Gazis et~al.\ (1963) average commute.
\section{Numerical testing}
Even though, in general, the Backus~(1962) average and the Gazis et~al.\ (1963) average do not commute, it is important to consider the extent of their noncommutativity.
We wish to enquire to what extent---in the context of a continuum-mechanics model and unavoidable measurement errors---the averages could be considered as approximately commutative.

To do so, we numerically examine two cases.
In one case, we begin---in the upper left-hand corner of Diagram~\ref{eq:CD}---with ten strongly anisotropic layers.
In the other case, we begin with ten weakly anisotropic layers.   

Elasticity parameters for the strongly anisotropic layers are derived by random variation from the H002 sanidine alkali feldspar given in Waeselmann et~al.\ (2016), but with the~$x_3$-axis perpendicular to the symmetry plane rather than the~$x_2$-axis, used by Waeselmann et~al.\ (2016).
These parameters are given in Table~\ref{tab:strong}.   

\begin{table}[H]
\caption{Ten strongly anisotropic monoclinic tensors.  The elasticity parameters are density-scaled; their units are $10^6~{\rm m}^2/{\rm s}^2$\,.}
\label{tab:strong}
{\footnotesize
\begin{tabular}{|c|c|c|c|c|c|c|c|c|c|c|c|c|c|}
\hline
layer&$c_{1111}$&$c_{1122}$&$c_{1133}$&$c_{1112}$&$c_{2222}$&$c_{2233}$&
$c_{2212}$&$c_{3333}$&$c_{3312}$&$c_{2323}$&$c_{2313}$&$c_{1313}$&$c_{1212}$\\
\hline
1&23.9&11.6&12.2&1.53&71.4&6.64&2.94&52.0&-2.89&8.00&-6.79&8.21&4.54\\
2&33.5&8.24&12.2&-0.98&66.9&5.65&2.02&82.3&-1.12&6.35&-5.16&17.4&7.36\\
3&33.2&9.79&16.9&0.57&62.1&6.19&3.81&83.4&-7.34&10.2&-2.33&16.6&4.72\\
4&38.1&8.33&12.2&1.51&55.0&4.87&3.11&56.8&-1.43&4.10&-0.20&8.25&11.2\\
5&37.4&11.5&14.4&-0.79&72.6&3.93&3.00&76.5&-6.07&9.58&-4.38&14.8&8.70\\
6&38.4&10.7&17.1&1.55&63.8&7.11&1.99&55.2&-0.98&9.66&-6.85&11.1&11.4\\
7&29.2&11.4&11.7&0.59&59.5&5.23&3.74&82.7&-3.81&10.1&-5.09&9.78&6.89\\
8&31.9&9.03&19.1&-0.07&71.6&4.18&1.98&70.4&-0.25&4.84&-0.33&8.21&10.9\\
9&37.5&10.5&19.4&0.37&76.7&5.02&3.57&76.7&-0.16&7.84&-1.62&13.8&10.7\\
10&36.0&9.65&18.9&-0.43&73.1&3.94&2.53&60.4&-7.20&5.44&-2.20&9.25&5.20\\
\hline
\end{tabular}}
\end{table}

Weakly anisotropic layers are derived from the strongly anisotropic ones by keeping $c_{1111}$ and $c_{2323}$\,, which are the two distinct elasticity parameters of isotropy, approximately the same as for the corresponding strongly anisotropic layers, and by varying other parameters away from isotropy.
These parameters are given in Table~\ref{tab:weak}.

\begin{table}[H]
\caption{Ten weakly anisotropic monoclinic tensors. The elasticity parameters are density-scaled; their units are $10^6~{\rm m}^2/{\rm s}^2$\,.}
\label{tab:weak}
{\footnotesize
\begin{tabular}{|c|c|c|c|c|c|c|c|c|c|c|c|c|c|}
\hline
layer&$c_{1111}$&$c_{1122}$&$c_{1133}$&$c_{1112}$&$c_{2222}$&$c_{2233}$&
$c_{2212}$&$c_{3333}$&$c_{3312}$&$c_{2323}$&$c_{2313}$&$c_{1313}$&$c_{1212}$\\
\hline
1&24&9&9&0.2&29&7&0.3&27&-0.3&8&-1&8.2&7\\
2&34&15&18&-0.1&38&14&0.2&39&-0.1&6&-1&7.5&6.5\\
3&33&12&14&0.06&37&10&0.4&38&-0.7&10&-0.5&12&8.5\\
4&38&20&22&0.15&40&15&0.3&41&-0.1&4&-0.2&5&6\\
5&37&14&16&-0.08&42&10&0.3&41&-0.6&10&-0.8&11&9\\
6&38&15&18&0.16&41&14&0.2&40&-0.1&10&-1&10.5&11\\
7&29&9.5&9.5&0.06&32&8&0.4&34&-0.4&10&-0.8&10&9\\
8&32&15&19.5&-0.01&36&13&0.2&36&-0.03&5&-0.3&6&6\\
9&38&16&20&0.04&43&14&0.4&42&-0.02&8&-0.4&9&9\\
10&36&18&23&-0.04&40&15&0.3&39&-0.7&5&-0.5&6&5\\
\hline
\end{tabular}}
\end{table}

Assuming that all layers have the same thickness, we use an arithmetic average for the Backus~(1962) averaging; for instance,
\begin{equation*}
\overline{c_{1212}}=\frac{1}{10}\sum_{i=1}^{10} c_{1212}^i\,.
\end{equation*}

The results of the clockwise and counterclockwise paths 
for the three elasticity parameters that differ from each other are given in
Table~\ref{tab:num}.
It appears that the averages nearly commute for the case of weak anisotropy.
Hence, we might conclude that the extent of noncommutativity is a function of the strength of anisotropy.

\begin{table}[H]
\caption{Comparison of numerical results.}
\label{tab:num}
\centerline{
\begin{tabular}{||c||c|c||c|c||c|c||}
\hline
&&&&&&\\[-12pt]
anisotropy&$c_{1212}^\circlearrowright$&$c_{1212}^\circlearrowleft$&$c_{1313}^\circlearrowright$&$c_{1313}^\circlearrowleft$&$c_{2323}^\circlearrowright$&$c_{2323}^\circlearrowleft$\\
\hline
strong&8.06&8.16&9.13&10.84&6.36&6.90\\
weak&7.70&7.70&7.88&7.87&6.82&6.81\\
\hline
\end{tabular}
}
\end{table}
\section{Strength of anisotropy}
To quantify the strength of anisotropy, we invoke the concept of distance in the space of elasticity tensors.
In particular, we consider the closest isotropic tensor according to the Frobenius norm, as formulated by Voigt (1910).
Examining one layer from the upper left-hand corner of Diagram~\ref{eq:CD}, we denote its weakly anisotropic tensor as $c^{\rm w}$ and its strongly anisotropic tensor as $c^{\,\rm s}$\,.

Using explicit expressions of Slawinski (2016), we find that the elasticity parameters of the closest isotropic tensor,~$c^{\rm iso_w}$\,, to $c^{\rm w}$ is $c^{\rm iso_w}_{1111}=25.52$ and $c^{\rm iso_w}_{2323}=8.307$\,.
The Frobenius distance from $c^{\rm w}$ to $c^{\rm iso_w}$ is $6.328$\,.

The closest isotropic tensor,~$c^{\rm iso_s}$, to $c^{\,\rm s}$ is $c^{\rm iso_s}_{1111}=39.08$ and $c^{\rm iso_s}_{2323}=11.94$\,.
The distance from $c^{\,\rm s}$ to $c^{\rm iso_s}$ is $49.16$\,.

Thus, as required, $c^{\,\rm s}$\,, which represents strong anisotropy,  is much further from isotropy than $c^{\rm w}$\,, which represents weak anisotropy.
\section{Discussion}
Dalton and Slawinski~(2016) show that---in general---the Backus~(1962) average and the Gazis et~al.\ (1963) average do not commute.
Herein,  using the the case of monoclinic and orthotropic symmetries, we numerically show that noncommutativity is a function of the strength of anisotropy.
For weak anisotropy, which is a common case of seismological studies, the averages appear to nearly commute.

In our future work, we will consider aspects of the approximation theory to rigorously examine  the commutativity issues between these averages.
Also, in such an examination, we will invoke advanced numerical methods. 
\section*{Acknowledgments}
This numerical examination was motivated by a fruitful discussion with Robert Sarracino.
The research was performed in the context of The Geomechanics Project
supported by Husky Energy. Also, this research was partially supported by the
Natural Sciences and Engineering Research Council of Canada, grant 238416-2013.
\section*{References}
\frenchspacing
\newcommand{\hd}{\par\noindent\hangindent=0.4in\hangafter=1}
\hd
Backus, G.E.,  Long-wave elastic anisotropy produced by horizontal layering,
{\it  J. Geophys. Res.\/}, {\bf 67}, 11, 4427--4440, 1962.
\setlength{\parskip}{4pt}
\hd 
Bos, L, D.R. Dalton, M.A. Slawinski and T. Stanoev,
On Backus average for generally anisotropic layers, {\it ar{X}iv\/} [{\bf physics.geo-ph}], 
1601.02967, 2016.
\hd
Dalton, D. R. and M.A. Slawinski, On commutativity of Backus average
and Gazis et~al. average, {\it ar{X}iv\/} [{\bf physics.geo-ph}], 1601.02969, 2016.
\hd
Gazis, D.C., I. Tadjbakhsh and R.A. Toupin, The elastic tensor of given symmetry nearest to an anisotropic elastic tensor, {\it Acta Crystallographica\/}, {\bf 16}, 9, 917--922, 1963.
\hd
Slawinski, M.A., {\it Waves and rays in seismology: Answers to unasked questions\/},
World Scientific, 2016.
\hd
Voigt, W., {\it Lehrbuch der Kristallphysik\/}, Teubner, Leipzig, 1910.
\hd
Waeselmann, N., J.M. Brown, R.J. Angel, N. Ross, J. Zhao and W. Kamensky, The
elastic tensor of monoclinic alkali feldspars, {\it American Mineralogist\/}, {\bf 101},
1228--1231, 2016.
\end{document}